# "Extreme" Microfluidics: Large-volumes and Complex Fluids

*Mehmet Toner - Massachusetts General Hospital Harvard-MIT*

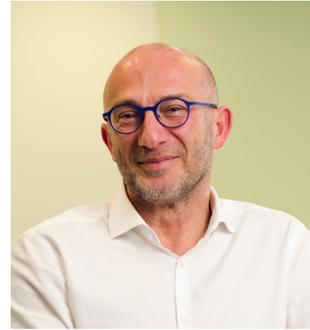

## Biography

Mehmet Toner is the Helen Andrus Benedict Professor of Biomedical Engineering at the Massachusetts General Hospital (MGH), Harvard Medical School, and Harvard-MIT Health Sciences & Technology. He is also the Director of the BioMicroElectroMechanical Systems Center at MGH, and Director of Research at the Shriners Hospital for Children Boston. Dr. Toner received BS degree from Istanbul Technical University and MS degree from the Massachusetts Institute of Technology (MIT), both in Mechanical Engineering. Subsequently he completed his PhD degree in medical engineering and medical physics at Harvard-MIT Division of Health Sciences and Technology in 1989. Dr. Toner published more than 350 research papers and his publications have been cited more than 45,000 with an h-index of 98. He also holds more than 50 patents and has co-founded multiple biotechnology start-ups. He is a "Fellow of the American Institute of Medical and Biological Engineering", "Fellow of the American Society of Mechanical Engineers", and "Fellow of the Society for Cryobiology." In 2012, he was given the "Luyet Medal" by the Society for Cryobiology. In 2013, he received the "H.R. Lissner Medal" from the American Society of Mechanical Engineering. In 2010, he was selected to the Board of Advisors of the National Science Foundation Engineering Directorate for a 3-year term. Dr. Toner is also a fellow of the "National Academy of Inventors" as well as a fellow of the "National Academy of Engineering."

## Abstract

Microfluidics gained prominence with the application of microelectromechanical systems (MEMS) to chemistry and biology in an attempt to benefit from the miniaturization of devices for handling of minute fluid samples under precisely controlled conditions. The field of microfluidics exploits the differences between micro- and macro-scale flows, for example, the absence of turbulence, electro-osmotic flow, surface and interfacial effects, capillary forces in order to develop scaled-down biochemical analytical processes.[1,2] The field also takes advantage of MEMS and silicon micromachining by integrating micro-sensors, micro-valves, and micro-pumps as well as physical, electrical, and optical detection schemes into microfluidics to develop the so-called "micro-total analysis systems (μTAS)" or "lab-on-a-chip" devices.[3,4] The use of silicon in the early years of miniaturization during 1980s was self-limiting to the growth of the field of microfluidics primarily due to its high cost and inaccessibility of manufacturing, and to some extent due to its incompatibility with light microscopy. After mid 1990s there has been a major leap in progress with the application of soft lithography using polydimethylsiloxane to engineer complex microfluidics devices as well as the ability to manufacture plastics at a micron scale resolution.[5,6]

In 2006, Dr. Whitesides described the field as: *"It [microfluidics] is the science and technology of systems that process or manipulate small ($10^{-9}$ to $10^{-18}$ litres) amounts of fluids, using channels with dimensions of tens of hundreds of micrometers"*.[1] On the other hand, the ability to process 'real world-sized' volumes ($10^{-4}$ to $10^{-1}$ litres) that would enable a wide range of industrial and medical applications has been a major challenge since the beginning of the field of microfluidics.[7] This begs the question whether it is possible to take advantage of microfluidic precision without the limitation on throughput required for large-volume processing? The challenge is compounded by the fact that physiological fluids are non-Newtonian, heterogeneous, and contain viscoelastic living cells that continuously responds to the smallest changes in their microenvironment. Starting in 2000s, we have initiated efforts to explore the use of microfluidics to process and manipulate volumes of physiological fluids that are five to ten orders of magnitude more than the amounts typically used in microfluidics.[8-12] Our efforts to advance the field of microfluidics to process large-volumes of fluids were counterintuitive and not anticipated by the conventional wisdom at the inception of the field. We showed that the new large-volume cell separation methods can be more than just a scale-up approach via parallelization of existing µTAS methods. We metaphorically called this "hooking garden hose to microfluidic chips."

In an early attempt, we developed a microfluidic chip to process milliliters volumes of whole blood (~$5 \times 10^{-3}$ liters) without any dilution to isolate extremely rare (1 in 1-10 billion blood cells) circulating tumor cells.[12] We used enabling aspects of physics of microfluidics to eliminate multiple steps used in bulk techniques to provide very uniform and gentle flow conditions to capture rare tumor cells in whole blood. We created a microfluidic platform that contains about 80,000 microposts that are distributed in such a manner to maximize the frequency of contact between cells and chemically modified microposts with an antibody to selectively bind tumor cells. The chip interrogates about $2 \times 10^{+6}$ cells·s$^{-1}$ or each micropost 20-30 cells·s$^{-1}$, which clearly brings the power of multiplexing and precise flow control inherent to microfluidics. Moreover, the maximum shear stress by a cell near the micropost surface was estimated to be 0.4 dyn·cm$^{-2}$, which is significantly lower than typical physiological shear stress experienced by cells, enabling sorting of live cells. Overall, this study has become a major driving force for the field and paved the way for subsequent studies exploring the use of microfluidics to manipulate complex bodily fluids.

Subsequently, we and others have also developed alternative approaches to use microfluidics for the isolation of rare cells from whole blood. We used micro-vortex generating device to ensure effective contacts of cells with antibody-coated surfaces to bind target cells using a simple geometry that is amenable for large-scale manufacturing.[13] We integrated ultra-porous forests of vertically aligned carbon nanotubes combining microporosity and nanoporosity to achieve simultaneous mechanical filtration and chemical bioparticle capture that enables simultaneous isolation of three different particle types ranging three orders of magnitude in size.[14] The use of ultra nanoporous posts improved particle interception compared to solid posts via the increase of direct interception and the reduction of near-surface hydrodynamic resistance. We also developed a microchip incorporating porous, fluid-permeable surfaces functionalized with cell-specific antibodies to capture a rare subpopulation of target cells with excellent selectivity and significantly enhanced throughput.[15] The efficiency of this microfluidic platform arises both from enhanced mass transport to porous surface and from enhanced cell-surface interactions that promote dynamic cell rolling adhesion with high specificity. These and other so-called 'positive' sorting microfluidic devices demonstrated the power of microfluidics for rare cell sorting from large-volumes of complex fluids.

More recently, we integrated multi-physics processing on a single microfluidic device to sort rare tumor cells from blood by "negative depletion," that is by removing tens of billions of blood cells without losing a few rare tumor cells.[16,17,18] Advantages of this approach for cancer detection are considerable in that the microfluidic device makes no *a priori* assumption about tumor cells and as such it applies to all cancers and all tumor cells in whole blood irrespective of size, phenotype, or other properties. After magnetic labeling of white blood cells using microbeads coated with antibodies, this microfluidic cell separation platform integrates three sequential microfluidic technologies to impart exquisite control over billions of particles, namely, (i) removal of smaller red blood cells and platelets from larger white blood cells and tumor cells using deterministic lateral displacement, (ii) the alignment of nucleated cells within a microfluidic channel using inertial focusing principles, and (iii) the deflection of magnetically tagged white blood cells into a waste channel while tumor cells sorted into a product channel. The chip processes 10-20 mL·h$^{-1}$ through ~1.5 million microscopic features, 256 individual deterministic lateral displacement arrays, and 4 parallel inertial focusing channels. More importantly, we use the power of microfluidic multiplexing to shorten the transit time in the chip to only less than a second so the cells are isolated unaltered for subsequent molecular analysis. The integration of multi-physics sorting of only a few rare cells at extremely high-throuput from blood containing over 50 billions of cells provides an excellent example of the enabling aspect of microfluidics physics and technology.

We also pioneered using inertial forces in microchannel flows for controlling cell and particle motion, which enabled extremely high-throughput manipulation and sorting of bioparticles.[19] As much as the framework in the field was that microfluidic systems operate at low Reynolds number in the absence of inertial forces, we showed that inertia is not only critical to low Reynolds number flows in microchannels but it can also be easily exploited to achieve exquisite control over manipulation of bioparticles and cells. We investigated the particle migration transverse to fluid streamlines in highly confined microchannel systems where the particle (cell) size approaches the channel dimensions. We used inertial forces in microfluidic systems in laminar flow at particle Reynolds number of $Re_p$~$O(1)$ to focus randomly distributed particles and cells continuously and at high rates to a single streamline. This effect was first reported experimentally for particles flowing through macroscale large pipes, the so-called "tubular pinch effect." However, no particle manipulation was accomplished at the macroscale flow mainly due to the difficulty in isolating particles from a large focused annulus in a tube flow. On the other hand, microscale effects in microchannel flows can be used to move particles across streamlines to predictable equilibrium positions within a flow and to differentially order particles of different sizes, continuously, at high throughputs, with no external forces, and also sort these particles by size and other physicochemical attributes.[20-23] We also used particle tracking analysis to probe the effect of inertial focusing of cells in whole blood (non-Newtonian) or diluted (Newtonian) blood.[24] We observed the inertial focusing behavior of white blood cells and tumor cells in whole blood despite steric interference from overwhelming number of red blood cells. Since the initial publication, we and others have explored inertial microfluidics to precisely manipulate and sort bioparticles.[25-26]

We also advanced the use of inertial microfluidics to several other untapped processes. We developed a microfluidic device that couples the phenomenon of inertial focusing and highly controllable hydraulic resistance micro-siphoning channels to achieve controlled bioparticle concentration.[27] Concentrating cells and bioparticles using centrifugation is a ubiquitous preparatory step in many biomedical applications and assays, and it also emerged as an immediate need in microfluidics, as devices that are capable of processing larger volumes of blood were producing dilute solutions of targeted cells. However, centrifugation suffers from long exposure of bioparticles to heightened centrifugal forces, formation of compacted pellet, and it is inherently variable and low yield process, especially when applied to larger volumes of samples. By taking the concept of repetitive fluid removal and refocusing to an extreme, a parallelized and serially integrated version of our microfluidic concentrator chip is able to reach a volume reduction factor >400 of a cellular suspension in a continuous manner. The microchip operates under continuous flow conditions, exposes cells passing through the device to uniform conditions, and has a transit time of only ~100 ms. More recently, we also developed a 'non-equilibrium' inertial separation method for ultra-high throughput and large-volume blood fractionation.[28] We devised a fully-parallelized microchip (104 devices) that can fractionate 400 mL of whole blood (containing ~2 trillion cells) within 3 hours to debulk red blood cells and platelets to produce white blood cell fractionate. The processing rate is >200 million cells·s$^{-1}$ resulting in fractionation of blood and bone marrow without clogging.[28] This was achieved because non-equilibrium inertial focusing in repeated units enables sorting cells by size without contacting surfaces. Despite the ultra-high throughput processing, the cells are alive and unaltered as the residence time on the chip is less than a second. More recently, we put forth the use of oscillatory flows in microchannels under inertial focusing conditions to effectively create an 'infinitely' long microchannel. This enabled us to focus particles that are sub-micron size in microchannels at very low pressure drops; hence enabling entirely new opportunities in a previously unexplored particle Reynolds number range (*i.e.* $Re_p \sim O(10^{-2})$).

In our quest to push the limits of inertial microfluidics, we also demonstrated inertia-elastic focusing of particles and cells in a previously entirely unexplored regime of channel Reynolds numbers of up to $Re \approx 10,000$ that was accessed through the use of a rigid microfluidic device.[29] We showed that on addition of micromolar concentrations of biopolymeric drug reducing agents, the resulting fluid viscoelasticity can be used to control the focal position of particles at Reynolds numbers up to $Re \approx 10,000$ in a single microchannel with corresponding flow rates and particle velocities up to 50 mL·min$^{-1}$ (or 3 L·h$^{-1}$) and 130 m·s$^{-1}$ (or 468 km·h$^{-1}$), respectively. We demonstrated that the focusing performance of the microfluidic device actually improves with increasing flow throughput up to $Re \sim O(10^3)$ and can be achieved when both elasticity and inertia are present (*i.e.*, $Wi \gg 1$ and $Re \gg 1$). We found that viscoelastic normal stresses drive the deterministic particle migration in this regime. We also note that our results represent an important improvement of three to four orders of magnitude in flow rate within only a single microchannel over previous studies of particle migration.

Finally, our recent efforts to use microfluidics have focused on collecting multicellular 'clusters' of tumor cells from whole blood.[30] Existing technologies for cell sorting are designed to isolate single tumor cells despite the fact that there is growing evidence that clusters of tumor cells are highly metastatic. We developed a microchip technology to capture tumor cell clusters from unprocessed blood. The cluster chip uses specialized bifurcating micro-traps under extremely low shear stress conditions at peak flow speed of ~70 $\mu m \cdot s^{-1}$, so that the clusters isolated independent of their deformability and the delicate cell-cell junctions are not damaged. The overall chip is multiplexed with 33,000 microscopic triangular structures forming 4,096 parallel trapping paths such that it can interrogate blood sample at an impressive overall volumetric flow rate of 2.5 $mL \cdot h^{-1}$ (~$4 \times 10^{-5}$ $lt \cdot s^{-1}$). We have advanced the cluster sorting to a continuous flow sorting approach using deterministic lateral displacement approach based on size and asymmetry.[31] Efficient sorting of living tumor clusters enables for the first time the detailed characterization of their role in metastasis.[32]

In summary, the use of microfluidics to process large-volumes of complex fluids has found broad interest in both academia and industry at an unprecedented rate due to its broad range of utility in medical and industrial applications. Microfluidics for the macro-world and related enabling technologies and fundamental physicochemical principles are likely to remain a highly active field for years to come.